\documentclass[fleqn,usenatbib,useAMS,twocolumn]{aastex63}
\usepackage{epstopdf}
\usepackage{epsfig,graphics,subfigure,psfrag,amsmath,amssymb,amscd,lineno}
\usepackage{url, hyperref, fancyhdr, multirow, aas_macros, makecell, verbatim} 

\usepackage{acronym}
\usepackage{CJK}
\usepackage{ulem, cancel} 

\def\apjl{Astrophys. J. Lett.}
\def\apjs{Astrophys. J. Suppl.}

\def\aap{Astron. Astrophys.}

\let\oldequation\equation
\let\oldendequation\endequation
\renewenvironment{equation}{\linenomathNonumbers\oldequation}{\oldendequation\endlinenomath}

\let\oldalign\align
\let\oldendalign\endalign
\renewenvironment{align}{\linenomathNonumbers\oldalign}{\oldendalign\endlinenomath}

\let\oldgather\gather
\let\oldendgather\endgather
\renewenvironment{gather}{\linenomathNonumbers\oldgather}{\oldendgather\endlinenomath}

\newcommand{\DOA}{Department of Astronomy, University of Science and Technology of China, Hefei, Anhui 230026, China. 
\href{Corresponding author.}{lianggui.zhu@ustc.edu.cn},
\href{Corresponding author.}{helei0831@mail.ustc.edu.cn},
\href{Corresponding author.}{wzhao7@ustc.edu.cn}   }

\newcommand{\SASS}{School of Astronomy and Space Sciences, University of Science and Technology of China, Hefei, Anhui 230026, China.}

\newcommand{\KIAA}{Kavli Institute for Astronomy and Astrophysics, Peking University, 
Beijing 100871, China.}
\newcommand{\DOAPKU}{Department of Astronomy, School of Physics, Peking University, Beijing 100871, China.
\href{Corresponding author.}{xian.chen@pku.edu.cn} }

\renewcommand\thetable{\Roman{table}}
\newcommand{\D}{\mathrm{d}}

\acrodef{GW}{gravitational-wave}
\acrodef{EM}{electromagnetic}
\acrodef{BH}{black hole}
\acrodef{BBH}{binary black hole}
\acrodef{BNS}{binary neutron star}
\acrodef{SNR}{signal-to-noise ratio}
\acrodef{AGN}{active galactic nucleus}
\acrodefplural{AGN}{active galactic nuclei}
\acrodef{CL}{confidence level}
\acrodef{3D}{three-dimensional}

\usepackage{orcid}

\begin{document}

\begin{CJK*}{UTF8}{gbsn} 

\title{Constraining the Fraction of LIGO/Virgo/KAGRA Binary Black Hole Merger Events Associated with Active Galactic Nucleus Flares }

\author{Liang-Gui Zhu ({\CJKfamily{gbsn}朱良贵})
\orcidlink{0000-0001-7688-6504} 
}
\affiliation{\DOA}
\affiliation{\SASS}

\author{Lei He ({\CJKfamily{gbsn}贺雷})
\orcidlink{0000-0001-7613-5815}
}
\affiliation{\DOA}
\affiliation{\SASS}

\author{Xian Chen ({\CJKfamily{gbsn}陈弦})
\orcidlink{0000-0003-3950-9317} 
}
\affiliation{\DOAPKU}
\affiliation{\KIAA}

\author{Wen Zhao ({\CJKfamily{gbsn}赵文})
\orcidlink{0000-0002-1330-2329}
}
\affiliation{\DOA}
\affiliation{\SASS}

\date{\today}
%
\begin{abstract}


The formation channels of binary black hole (BBH) mergers detected by the LIGO/Virgo/KAGRA (LVK) network remain uncertain. 
While BBH mergers occurring inside active galactic nucleus (AGN) disks may interact with surrounding gas 
and generate observable optical flares. We test this scenario by quantifying the spatial and temporal correlation 
between BBH events in GWTC-4.0 and AGN flares identified from six years of the Zwicky Transient Facility (ZTF) DR23 data. 
Using 80 BBH mergers selected for adequate localization, redshift reach, observing-epoch overlap, and ZTF sky coverage, 
we construct a likelihood for the flare-associated fraction, $f_{\rm flare}$, that combines each event's 3D localization 
with a locally estimated flare number density derived from a 3D Voronoi tessellation, while explicitly accounting for 
survey boundaries and incomplete catalog coverage. Adopting a 200-day post-merger time window for potential counterparts, 
we infer $f_{\rm flare} = 0.07_{-0.05}^{+0.24}$ (90\% confidence level). This non-zero maximum-likelihood value is driven 
primarily by GW190412, for which a single flare candidate (J143041.67+355703.8) is consistent in both time and spatial position. 
The candidate's light curve is limited to two data points during its peak, 
so it remains classified only as a candidate AGN flare. 
Excluding GW190412 yields results consistent with no association and an upper limit of $f_{\rm flare} < 0.17$ 
at 90\% confidence level. The intrinsic properties of GW190412 and the characteristics of the candidate host AGN 
are broadly consistent with theoretical expectations for the AGN-disk formation channel, motivating continued, 
targeted electromagnetic follow-up of well-localized and highly asymmetric BBH mergers in current 
and upcoming time-domain surveys.

\end{abstract}


\section{Introduction}     \label{sec:introduction}

The LIGO/Virgo/KAGRA (LVK) ground-based \ac{GW} detector network has officially released 
version 4.0 of Gravitational-Wave Transient Catalog (GWTC-4.0) \citep{2025ApJ...995L..18A, 2025arXiv250818082T}. 
This catalog presents $176$ compact binary coalescence events with parameter estimates 
(false alarm rate $< 1 ~\!{\rm yr}^{-1}$),
covering observations from the first observing run ($O1$)
through the first part of the fourth observing run \citep[$O4$a;][]{2019PhRvX...9c1040A, 
2021PhRvX..11b1053A, 2023PhRvX..13d1039A, 2024PhRvD.109b2001A}. 
More than $90\%$ of the reported events originate from \ac{BBH} mergers. 
The properties of these \ac{BBH} events have substantially advanced 
our understanding of stellar-mass \ac{BH} and \ac{BBH} populations 
\citep[e.g.,][]{2016PhRvL.116x1102A, 2019ApJ...882L..24A, 2020ApJ...900L..13A, 2020PhRvD.102d3015A, 
2020PhRvL.125j1102A, 2025ApJ...993L..25A, 2025ApJ...993L..21A, 2025arXiv250818083T}. 
However, the origins of the observed \acp{BBH} remain highly uncertain. 

Several formation channels have been proposed, including 
the evolution of primordial black holes \citep{1974Natur.248...30H, 2025LRR....28....1B}, 
remnants of Population III stars \citep{2010MNRAS.403...45S, 2017MNRAS.470..898H}, 
and isolated binary evolution in galactic fields \citep{1998ApJ...506..780B, 2002ApJ...572..407B, dominik12, 
2014LRR....17....3P, 2016ApJ...824L..10W, 2020MNRAS.499.5941D}. 
Alternatively, BBHs may form dynamically through hierarchical mergers in dense environments, such as 
young stellar clusters \citep{ziosi14, 2017MNRAS.467..524B, decarlo19}, 
globular clusters \citep{2002MNRAS.330..232C, 2003ApJ...599.1260C, 2015PhRvL.115e1101R, 2016PhRvD..93h4029R, 2018PhRvL.121p1103F}, 
nuclear star clusters \citep{antonini16, antonini19}, 
and \ac{AGN} disks \citep{mckernan12, bartos17gas, yang19, 2020ApJ...898...25T}. 
Each proposed channel is theoretically capable of producing a subset of the detected BBH events
\citep{2021MNRAS.504L..28K, 2021MNRAS.502.2049L, 2024PhRvL.133e1401L, 2025ApJ...981..177L}. 
However, the predicted distributions of intrinsic BBH properties, 
exhibit significant overlap across channels, despite their distinct characteristic features. 
This degeneracy, coupled with the insufficient precision of GW source localization (for identifying the unique host of a GW source), 
poses a significant challenge for identifying the formation pathways of individual BBH events \citep[e.g.,][]{2025ApJ...993L..30L, 
2025arXiv250801135T, 2025arXiv250717551L, 2025arXiv250809965D, 2025arXiv250813412D, 2025arXiv250908298L, 2025arXiv251014363P}. 
Consequently, this challenge necessitates relying on population-level statistical methods to infer BBH formation channels, 
namely, estimating the fractional contribution of each proposed channel to the overall observed BBH population. 

Currently proposed methods for such inference include: 
i) hierarchical Bayesian inference \citep{2004AIPC..735..195L, 2018ApJ...863L..41F, 2019MNRAS.486.1086M, 2019PASA...36...10T}, 
ii) spatial correlation tests between BBHs and AGNs \citep{2008APh....29..299B, bartos17method, veronesi22, veronesi23obs, veronesi24obs}, and 
iii) spatial correlation tests between BBHs and \ac{EM} transients \citep[e.g., flares,][]{veronesi24flares} 
\citep[also see][with similar principles but a different statistical framework]{2021ApJ...914L..34P}. 
The first method quantifies the consistency between the observed posterior distributions of BBH source parameters 
\citep[e.g., component masses, mass ratios, spins, effective spins, and orbital 
eccentricities,][]{zevin21, Santoliquido21, 2021MNRAS.505..339M, 2021hgwa.bookE..16M, 2023PhRvD.107f3007L, 2025arXiv251203152B} 
and the predicted distributions from theoretical formation channels. 
Its key advantage lies in not relying on additional observational data, 
but the results are highly sensitive to the accuracy of the theoretical models themselves 
\citep{2019ApJ...882L..24A, gayathri21, wang21bbh-ang-mass-spin, 2022ApJ...941L..39W, 2024ApJ...977...67L, 2025ApJ...981..177L, 2025arXiv250818083T}.

The second method is specifically designed to test the AGN formation channel for BBHs \citep{bartos17method}. 
AGN disks are considered a favorable environment for producing LVK BBH events due to several key characteristics 
\citep{bartos17gas, yang19, 2020ApJ...898...25T, gayathri21, 2021ApJ...910...94C}: 
a number density of compact objects far exceeding that in galactic nuclear star clusters; 
dense gas that can fuel BH growth via accretion; and 
gas-induced dynamical friction, which promotes the efficient formation of bound multi-BH systems, 
enhances merger rates, and leads to spontaneous spin alignment, 
even for BBHs with 
distinctive properties such as larger masses, smaller mass ratios, 
higher pre-merger spins, higher effective inspiral spin magnitudes, and higher eccentricities 
\citep{yang19mass-spin, 2020ApJ...899...26T, 2021ApJ...908..194T, 
samsing22, 2025ApJ...987...65L, 2025arXiv250923897L}. Using this method, \citet{Zhu:2025rdv} reported 
preliminary evidence for a statistical association between LVK \ac{GW} events and AGNs. 

The third method extends the second by further restricting the spatial correlation targets to 
AGNs that exhibit temporally coincident \ac{EM} transients \citep{veronesi24flares}. 
Within the framework of general relativity, BBH mergers in vacuum do not produce \ac{EM} emission. 
However, if a \ac{BBH} merger occurs within an AGN disk, the interaction of its component black holes--and 
the remnant black hole--with the surrounding gas before and after the merger can potentially generate 
observable \ac{EM} counterparts \citep{2019ApJ...884L..50M, 2021ApJ...916..111K, 2021ApJ...916L..17W, 2024MNRAS.527.6076R, 2024ApJ...961..206C}. 
The confirmed spatial correlation between a fraction of BBH mergers and such AGN \ac{EM} transients 
can be interpreted as evidence that those BBHs formed in AGN disks. 

\citet{veronesi24flares} first applied the third method to statistically test the spatial correlation between 
BBH events detected during the third LVK observing run 
and candidate AGN flares identified from the Zwicky Transient Facility (ZTF) data by \citet{graham23}, 
finding no evidence for an association 
(a result also supported by \citet{2025arXiv251020767C} using a different framework). 
However, observations from other studies provide some support for the possibility that a subset of BBH mergers 
may be accompanied by flare counterparts. For instance, \citet{graham20flare} reported a high-confidence AGN flare counterpart 
associated with GW190521 
(see \citet{2021ApJ...914L..34P} for a detailed analysis), 
and later works \citep{2024PhRvD.110l3029C, 
2025arXiv250800291Z, 2025ApJ...990..154H, 2025arXiv251105144H} 
have identified other candidate AGN flares potentially correlated with other BBH mergers. 
The non-detection in \citet{veronesi24flares} is likely attributable to the low completeness of the flare catalog used. 

In this work, we employ updated observational data and an improved statistical framework to analyze 
the spatial correlation between LVK \ac{BBH} mergers and candidate flares, 
thereby inferring the fraction of detected BBH merger events with associated flare counterparts. 
The enhanced completeness of the updated dataset increases the probability of 
capturing genuine \ac{EM} counterparts, while the refined method more effectively 
accounts for spatial anisotropy in the flare catalog, reducing potential systematic biases. 

This paper is organized as follows. 
Section~\ref{sec:data} describes the data, 
Section~\ref{sec:method} introduces the statistical framework, 
Section~\ref{sec:results} presents the findings, 
and Section~\ref{sec:discussions} provides the discussion. 
We adopt a flat-$\Lambda$CDM cosmology 
$(H_0 = 67.8~\mathrm{km \cdot s^{-1}\! \cdot Mpc^{-1}},~ \Omega_M = 0.308)$ 
\citep{2016A&A...594A..13P} for redshift-distance conversion throughout this work.

\section{Data}   \label{sec:data}

The data used in this work consist of skymaps of \ac{BBH} \ac{GW} sources and \ac{AGN} flare catalogs. 
The skymaps of \ac{BBH} sources are obtained from the results of GWTC-4.0 of the LVK Collaboration, 
including the officially published \ac{BBH} events from the first observing run to the first part of the fourth observing run ($O1-O4$a)
\citep{2019PhRvX...9c1040A, 2021PhRvX..11b1053A, 2023PhRvX..13d1039A, 2024PhRvD.109b2001A, 
2025arXiv250818082T} \footnote{GWTC-4.0 is available at:
\url{https://gwosc.org/eventapi/html/GWTC/} }. 
These skymaps provide us with the probability density distribution function 
of the \ac{3D} localization of each \ac{BBH} source. 

The AGN flare catalog used in this work is adopted from \citet{2026ApJS..282...13H}. 
This catalog was constructed by searching for six years (March 2018 to October 2024) 
of data from the ZTF \citep{2019PASP..131g8001G, 2019PASP..131f8003B} 
\footnote{ZTF Data Release 23 is available at: \url{https://www.ztf.caltech.edu/ztf-public-releases.html} }. 
\citet{2026ApJS..282...13H} provides two versions: 
a refined \ac{AGN} flare catalog (AGNFRC) containing $\sim 2,000$ high-confidence flares, and 
a coarse \ac{AGN} flare catalog (AGNFCC) containing  $\sim 28,000$ candidate flares. 
Both were identified using Bayesian blocks and Gaussian processes, with the AGNFRC being a subset of the AGNFCC. 
Given the still limited understanding of the observational characteristics of genuine BBH merger-associated flares, 
we utilize the more complete AGNFCC to search for candidate counterparts in this work, 
after removing flares that have been identified as TDEs, SNe, and Blazars from the catalog. 
The used AGNFCC covers approximately $55\%$ of the sky. 

To enhance computational efficiency, we applied a selection to the BBH events, 
retaining only those satisfying the following criteria for our analysis: 
i) a spatial localization error volume $\Delta V_{\rm c}$ smaller than $10^{10}~\!{\rm Mpc}^3$, since poorer localization would cause the statistical signal of 
its unique flare counterpart (if it exists) to be more easily drowned out by numerous flare candidates; 
ii) a fraction of the 90\% \ac{CL} localization error volume covered by the flare catalog greater than $0.2$, 
because a lower coverage fraction increases the probability that the true flare counterpart of the BBH event was missed observationally; 
iii) the entire localization error volume lying within $z < 1.5$, a redshift range containing approximately 90\% of the cataloged flares; and 
iv) a merger time falling within the observational epoch of the ZTF data. 
In this work, we utilized a total of 80 BBHs from GWTC-4.0 that met these criteria. 
Our selection is independent of BBH intrinsic properties (and hence of formation channels), 
and the resulting subsample (representing $\sim 50\%$ of the total) remains representative 
of the full population regarding potential flare associations.

\section{Methodology}  \label{sec:method}

\subsection{Statistical framework}  \label{sec:method_framework}

In this work, we adopt the statistical framework presented in \citet{Zhu:2025rdv} to test 
the spatial correlation between the LVK BBH events and ZTF flares. 
This framework was originally introduced by \citet{bartos17method} to assess spatial correlations between GWs and AGNs. 
It was later modified by \citet{veronesi23obs} and \citet{veronesi24flares} 
to incorporate skymap information from GW events more effectively. 
\citet{Zhu:2025rdv} further extended the method by implementing a \emph{Voronoi tessellation} approach 
to address the significant anisotropies in the AGN catalog. 
Given that the flare catalog used here also exhibits significant anisotropy, 
the framework of \citet{Zhu:2025rdv} is particularly well-suited for our analysis. 

Let $f_{\rm flare}$ denote the fraction of BBH merger events consistent with the observed \ac{EM} flares. 
For a dataset consisting of $N$ BBH events, the likelihood can be expressed as 
\begin{align}  \label{eq:likeli_tot}
\mathcal{L}(f_{\rm flare}) =  \prod_{i=1}^{N}  \bigg[ & 0.9 \cdot f_{{\rm c},i} \cdot f_{\rm flare} \cdot \mathcal{S}_i \nonumber \\
 + & \big( 1 - 0.9 \cdot f_{{\rm c},i} \cdot f_{\rm flare} \big) \cdot \mathcal{B}_i    \bigg],
\end{align}
where $f_{{\rm c},i}$ represents the coverage fraction of the flare catalog for the localization error volume of the $i$-th GW event, 
and the coefficient $0.9$ comes from the \ac{CL} employed in the skymaps of BBH events. 
The functions $\mathcal{S}_i$ and $\mathcal{B}_i$ represent the ``signal probability'' and ``background probability'', respectively. 
These probabilities are constructed on the following principle: the statistical expectation of
$\mathcal{S}_i$ is greater than that of $\mathcal{B}_i$ if the $i$-th BBH event is consistent with an observed flare and vice versa. 
A theoretical proof provided in the appendix of \citet{Zhu:2025rdv} demonstrates that 
an unbiased estimate of $f_{\rm flare}$ can be achieved using the likelihood form of Eq. (\ref{eq:likeli_tot}). 

The signal probability is defined as 
\begin{equation}  \label{eq:prob_Si}
\mathcal{S}_i = \sum_{j=1}^{N_{{\rm flare}}} \frac{ p_i(\mathbf{x}_j) }{n_{\rm flare}(\mathbf{x}_j) }, 
\end{equation}
where $\mathbf{x}_j$ denotes the \ac{3D} position of the $j$-th candidate flare, 
$p_i(\mathbf{x}_j)$ represents the spatial localization probability density of the $i$-th BBH event at position $\mathbf{x}_j$, 
$n_{\rm flare}(\mathbf{x}_j)$ represents the spatial number density of the $j$-th candidate flare, 
and $N_{{\rm flare}}$ is the total number of candidate flares of the BBH event. 
The search of the candidate flares requires spatial and temporal coincidence with the BBH event: 
i) located within the 90\% \ac{CL} error volume of the BBH event; 
ii) the rise start and peak times of the flare both occur within 200 days after the BBH merger.
The experimental search window was set based on the \ac{EM} flare production model 
of the Ram-pressure Stripping \citep{2019ApJ...884L..50M}, 
when the time delay of the flare counterpart for a BBH merger GW event falls within this window, 
the flare is more likely to be observable.
This search window aligns with previous researches \citep{graham20flare, graham23, veronesi24flares}. 

To derive $n_{\rm flare}$ as a function of $\mathbf{x}_j$, we follow \citet{Zhu:2025rdv} 
and apply a \ac{3D} first-order Voronoi tessellation method 
\citep{voronoi1908nouvelles, 1992stca.book.....O, brakke2005statistics, 2009JSP...134..185L, 2021inas.book...57V}. 
Since the Voronoi tessellation is suitable for large samples, 
we first select all flares that are temporally consistent with each utilized BBH event from the total flare catalog 
according to the criteria for searching for candidate flare counterparts of BBHs outlined in \citet{graham23}: 
i) the Gaussian-rise start time ($t_0-1.8 t_{\rm g}$) and the peak time $t_0$ of the flare occurs within 200 days after the BBH merger, 
ii) the Gaussian rise timescale $t_{\rm g}$ of the flare is $t_{\rm g} \leq 100$ days, and 
iii) the exponential decay timescale $t_{\rm e}$ of the flare is $t_{\rm e} \leq 200$ days. 
These flares form a sub-catalog, we then use the Voronoi tessellation method to partition the comoving volume covered by 
the flare sub-catalog into polyhedral cells within a Cartesian coordinate system, 
with each cell containing exactly one flare. 
The flare density is equal to the reciprocal of the volume of the Voronoi polyhedron cell, 
i.e., $n_{\rm flare}(\mathbf{x}_j) = 1 / V_{\rm cell}(\mathbf{x}_j)$. 

The background probability $\mathcal{B}_i$ is defined as the expected value of $\mathcal{S}_i$ 
when the BBH event is unrelated to all observed candidate flares.
Because $\mathcal{S}_i$ merely performs a summation over the probability density distribution of BBH source spatial localization 
using observed flare samples, and flare catalogs are often incomplete--potentially covering only a portion of 
the localization error volume of the BBH event--the definition of $\mathcal{B}_i$ must account for this limitation. 
The background probability can be expressed as 
\begin{align}  \label{eq:prob_Bi}
\mathcal{B}_i = 0.9 \cdot f_{{\rm c},i} . 
\end{align}
The coverage fraction can be calculated by
\begin{align}  \label{eq:f_cover}
f_{{\rm c},i} = \frac{1}{\mathcal{N}_{\rm CL}} \iiint\limits_{\Delta V_{\rm c, flare}} p_i(\alpha, \delta, D_{\rm c}) |J|\D\alpha \D\delta \D D_{\rm c}, 
\end{align}
where $(\alpha, \delta)$ are celestial coordinates, $D_{\rm c}$ is the comoving distance, 
$p_i(\alpha, \delta, D_{\rm c})$ represents the spatial probability density distribution of the $i$-th BBH source localization, 
$|J| \equiv D_{\rm c}^2 \cos\delta$ is the Jacobian determinant for spherical-to-cartesian coordinate transformation, 
and $\mathcal{N}_{\rm CL} = 0.9$ is the normalization factor that accounts for the 90\% \ac{CL} localization error volume. 
The integration domain $\Delta V_{\rm c, flare}$ is the portion of the 90\% \ac{CL} localization volume of 
the $i$-th BBH event that is covered by the union of the Voronoi cells associated with all flares used in the analysis.

\subsection{Accounting for the boundary effects}  \label{sec:method_boundary}

An examination of the likelihood formula in Eq. (\ref{eq:likeli_tot}) reveals a complete degeneracy 
between the parameters $f_{{\rm c},i}$ and $f_{\rm flare}$. Consequently, the reliability of estimating 
$f_{\rm flare}$ depends directly on the accuracy of the $\{f_{{\rm c},i} \}$ calculations. 
For the majority of GW events in our sample--those whose localization error volume is only partially covered by 
the flare catalog--the calculation of $f_{{\rm c},i}$ is highly sensitive to the adopted sky-coverage boundaries. 
Therefore, it is essential to rigorously account for boundary effects in the data analysis. 

To accurately account for boundary effects, we performed two data-processing steps. 
First, we employed the \textsf{alphashape} algorithm to determine the sky-coverage boundaries of the flare distribution 
from the full flare catalog, excluding isolated flares and small-scale clusters located outside these boundaries. 
When calculating the volumes of the polyhedral cells (occupied by flares) using the Voronoi tessellation method, 
we removed vertices lying outside the sky boundaries (including those with redshifts exceeding the maximum threshold of $z = 1.5$). 
This procedure ensures that all Voronoi polyhedral cells reside within the sky boundaries and possess finite volumes. 
However, at this stage, the outer surfaces of the resulting polyhedral tessellations (formed by the retained flares) 
do not coincide with the original sky boundaries. 
This discrepancy becomes more pronounced in regions of lower flare spatial density. 
The volumes enclosed by these polyhedral tessellation surfaces define the precise integration region, 
$\Delta V_{\rm c, flare}$, for calculating the coverage fraction $f_{{\rm c},i}$ in Eq. (\ref{eq:f_cover}). 

Second, to precisely determine the volumes occupied by the polyhedral tessellations, 
we discretize the total volume (enclosed by the sky boundaries) into numerous minute cubic elements within 
the comoving-distance space. Each element is labelled by the coordinates of its center point. 
The volume of each Voronoi polyhedral cell (associated with a flare) is then determined by employing the 
\textsf{scipy.spatial.Delaunay} algorithm to select all cubic elements whose center points lie inside that cell. 
The union of these selected elements defines the integration region for calculating $f_{{\rm c},i}$. 
Note that  the computational precision of $f_{{\rm c},i}$ depends critically on the size of the cubic elements, 
which must be chosen to be much smaller than the smallest Voronoi cell volume in the sample. 

\section{Results}   \label{sec:results}

\subsection{Indications of a non-zero $f_{\rm flare}$}   \label{sec:result_f_flare}

\begin{figure}[tbp] 
 \centering
 \includegraphics[width=0.47\textwidth]{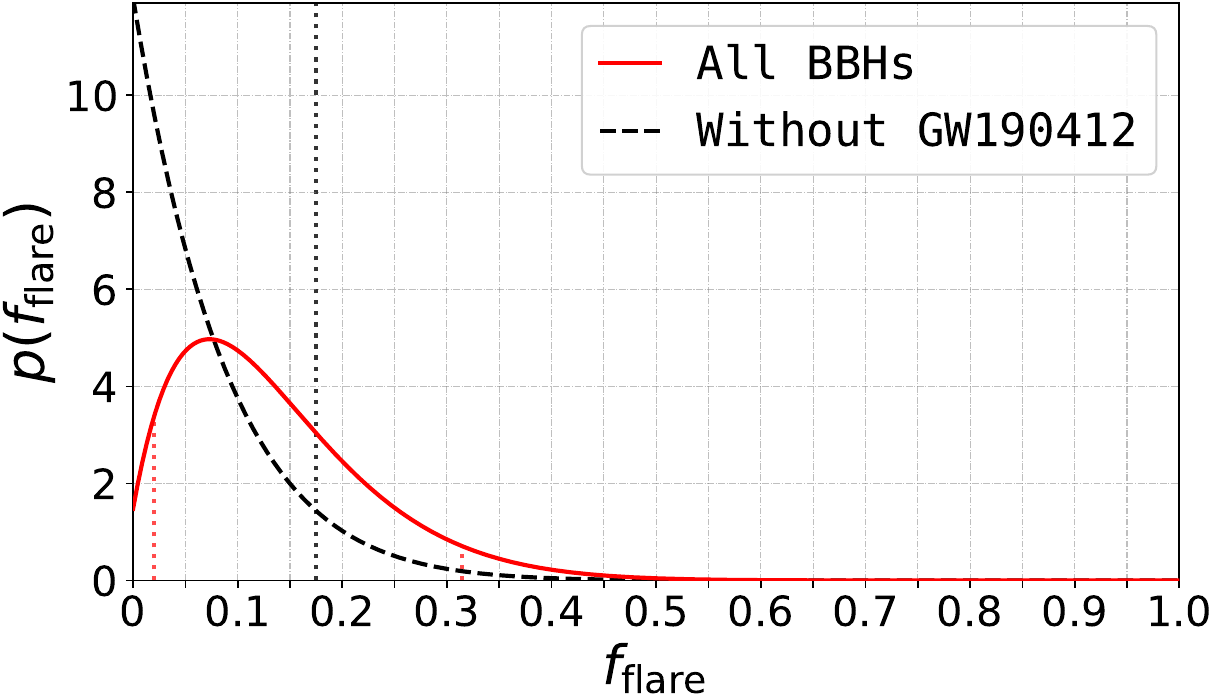}
 \caption{Probability density distributions of $f_{\rm flare}$ derived from the LVK GWTC-4.0 BBH events and the ZTF AGN candidate flares. 
 The red solid and black dashed curves represent the results derived from the all BBH events and 
 the set without \ac{GW}190412, respectively. The vertical red and black dotted lines indicate 
 the error interval and the upper limit at the 90\% \ac{CL}, correspondingly. 
 }
 \label{fig:f_flare_PDF}
\end{figure}

We first employ a 200-day post-merger search window and 
($t_{\rm g} \leq 100 \!~{\rm days},~ t_{\rm e} \leq 200 \!~{\rm days}$) condition to identify all flares 
temporally coincident with LVK BBH events from AGNFCC. 
The fraction of BBHs associated with flares, $f_{\rm flare}$, is estimated using Eq. (\ref{eq:likeli_tot}), 
and the resulting estimates are presented in Figure \ref{fig:f_flare_PDF}. 

The key distinction between our estimates and those from previous studies \citep[e.g.,][]{veronesi24flares, 2025arXiv251020767C} is 
the detection of a non-zero best-fit value for $f_{\rm flare}$. 
We estimate $f_{\rm flare} = 0.07_{-0.05}^{+0.24}$ (maximum-likelihood value with symmetric 90\% \acp{CL}, 
see the red solid curve in Figure \ref{fig:f_flare_PDF}). 
This result provides the first statistically derived preliminary evidence that a fraction of \ac{BBH} merger events 
are accompanied by observable \ac{EM} flares, representing a significant step forward since 
the candidate flare search for the GW190521 event was reported in \citet{graham20flare}. 

Before drawing a definitive conclusion, we investigated the origin of the signal for this non-zero $f_{\rm flare}$ evidence. 
We examined the signal probability $\mathcal{S}_i$ for each individual BBH event, 
as a higher $\mathcal{S}_i$ provides stronger support for $f_{\rm flare} > 0$. 
We found that GW190412 has an exceptionally high signal probability of $\mathcal{S}_i > 10$, 
which far exceeds those of the other BBH events. After removing it, 
the estimated $f_{\rm flare}$ becomes consistent with zero (black dashed curve in Figure \ref{fig:f_flare_PDF}), 
with an upper limit of $0.17$ at the $90\%$ \ac{CL}, 
in agreement with the result from \citet{veronesi24flares}. 
This evidence of $f_{\rm flare} > 0$ is primarily driven by \ac{GW}190412. 

\begin{figure}[tbp]
 \centering
 \includegraphics[width=0.45\textwidth]{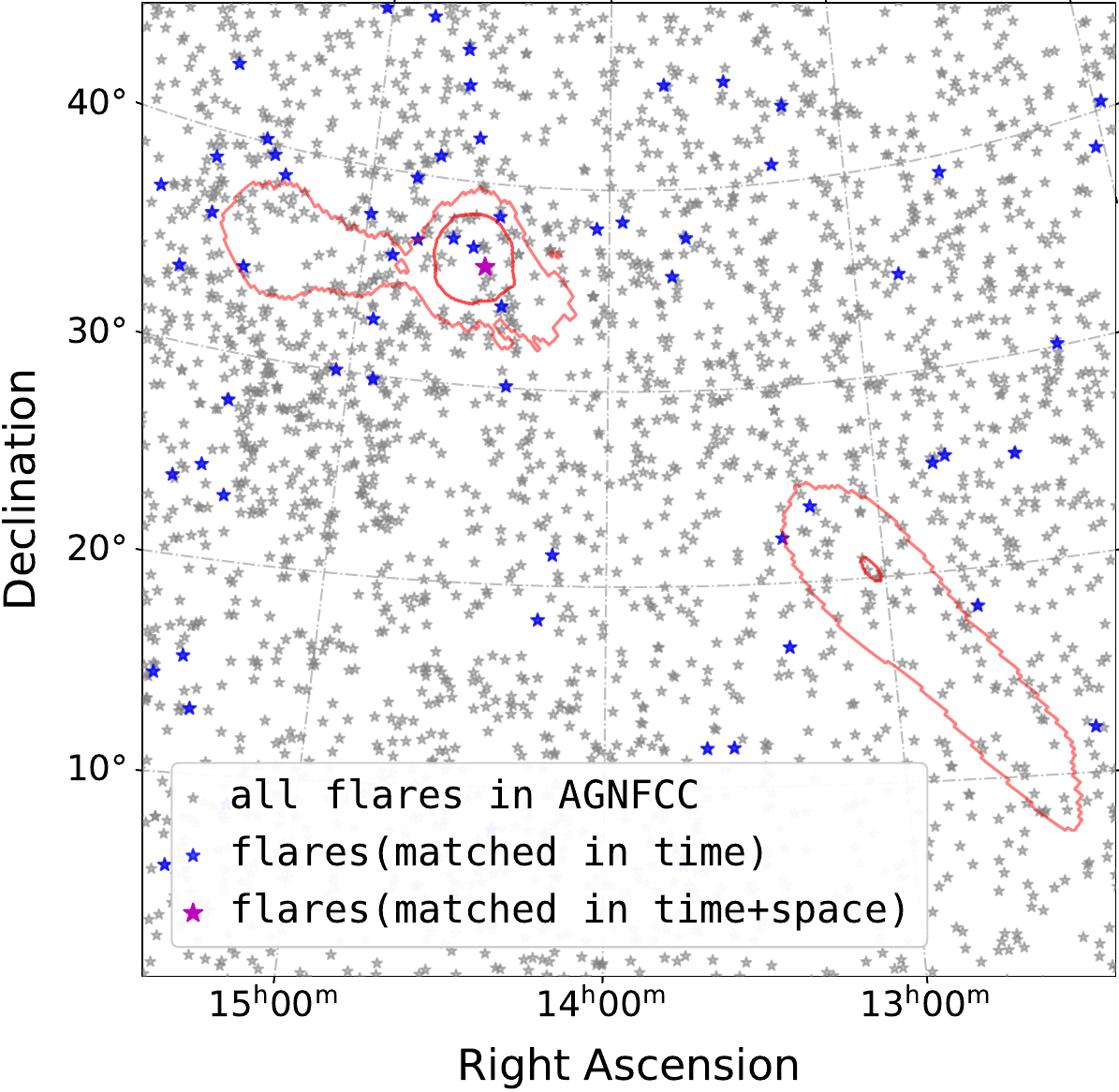}
 \caption{Skymap of \ac{GW}190412 event and candidate flares. 
 The red contours represent the 50\% and 90\% \ac{CL} sky localization regions, respectively. 
 The gray, blue, and purple star points denote all flares in AGNFCC, 
 those consistent with \ac{GW}190412 in time, 
 and those consistent in both time and spatial localization (sky position+distance), correspondingly. 
}
 \label{fig:gw_skymap}
\end{figure}

However, this does not imply that the preliminary evidence we obtained for $f_{\rm flare} > 0$ is unreliable. 
Figure \ref{fig:gw_skymap} presents the skymap of \ac{GW}190412 localization and 
the distribution of flares from AGNFCC. The gray points represent all AGNFCC flares, 
the blue points mark flares temporally coincident with the \ac{GW}190412 merger, and 
the purple point identifies the sole candidate: J143041.67+355703.8, consistent with the \ac{GW}190412 merger 
in both time and spatial position. 
A number of gray points and several blue points fall within 
the sky localization region of \ac{GW}190412, indicating that the survey coverage of AGNFCC was 
relatively complete across this sky area during the merger epoch. 
This suggests that the high signal probability derived from the J143041.67+355703.8 candidate flare 
is unlikely to be merely an artifact of residing in an underdense region (or ``cavity'') of the flare distribution, 
where a low local flare density $n_{\rm flare}$ would lead to an artificially high signal probability $\mathcal{S}_i$ 
(see Eq. (\ref{eq:prob_Si})). 

Additionally, among the $80$ BBH events used in this work, we find that $12$ BBH events have 
localization error volumes comparable to or smaller than that of \ac{GW}190412 (within a factor of 3). 
Within this subset, $8$ events exhibit a signal probability of zero, 
and the average ratio of signal probability to background probability is close to unity. 
This indicates that although the signal probability of \ac{GW}190412 itself is an outlier, 
the overall distribution of signal probabilities for events with similarly precise localizations 
is statistically consistent with a balanced population.

\subsection{Light curve and host AGN properties of the candidate flare counterpart for \ac{GW}190412}   \label{sec:result_lightcurve}

\begin{figure}[tbp]
 \centering
 \includegraphics[width=0.45\textwidth]{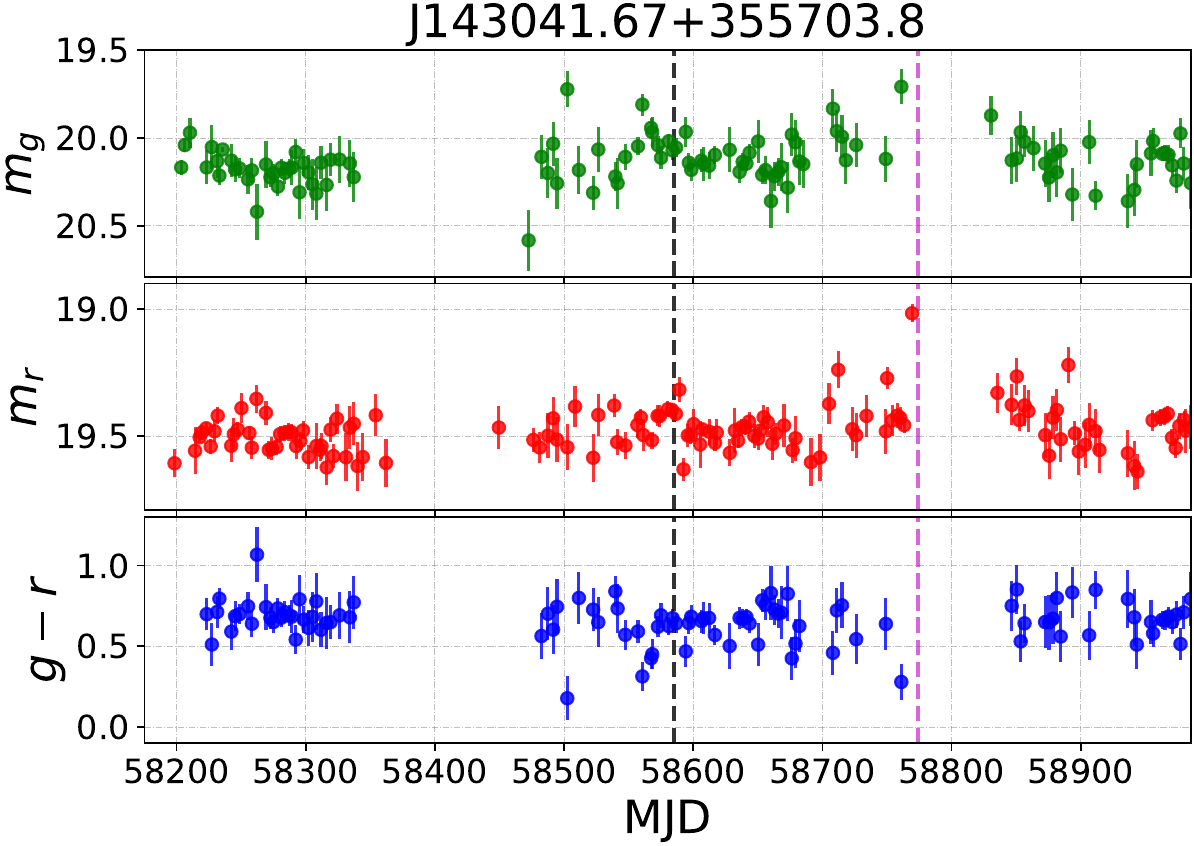}
 \caption{The ZTF $g$- and $r$-band light curves for the J143041.67+355703.8 candidate flare associated with GW190412. 
 The green and red points (with error bars) show the $g$- and $r$-band apparent magnitudes, 
 respectively; the blue points denote the $g-r$ color. 
 The vertical black and magenta dashed lines mark the merger time of GW190412 and the candidate flare's peak time, respectively. 
}
 \label{fig:flare_lightcurve}
\end{figure}

The light curve of the J143041.67+355703.8 candidate flare of \ac{GW}190412 is shown in Figure \ref{fig:flare_lightcurve}. 
When fitting the light curve using an empirical form 
\begin{equation}  
y(t) = \left\{  \begin{array}{ll}
 r_0 + A \exp \!\left[ -\frac{(t - t_0)^2}{2t_{\rm g}^2} \right],    & t \leq t_0    \nonumber \\ [6pt]
 r_0 + A \exp \!\left[ -\frac{(t - t_0)}{t_{\rm e}}      \right],         & t > t_0,      \nonumber
 \end{array} \right.
\end{equation}
the best-fit Gaussian rise and exponential decay timescales are $t_{\rm g} = 3$ days and $t_{\rm e} = 24$ days, respectively, 
with the fitted reference flare peak time $t_0$, amplitude $A$ and background flux $r_0$. 
The time delay between the merger of the \ac{GW}190412 event and 
the peak of the J143041.67+355703.8 candidate flare is about 189 days. 
As can be seen from Figure \ref{fig:flare_lightcurve}, there is unfortunately only one observational data point 
in each of the $g$ and $r$ bands around the peak time of the J143041.67+355703.8 candidate flare, corresponding to 
a magnitude variation of $\sim 0.5 ~\!{\rm mag}$. Although the probability of J143041.67+355703.8 being a genuine flare, 
estimated using the test statistic for flare searches employed by \citet{2026ApJS..282...13H}, 
reaches $p_{\mathrm{flare}} > 0.99$, it is still insufficient to 
draw a strong conclusion comparable to that in \citet{graham20flare}. 

We specifically note that the peak data point of the J143041.67+355703.8 candidate flare corresponds to 
the final observation in a contiguous ZTF monitoring block for this source. 
A manual inspection of hundreds of ZTF light curves of flares in the AGNFCC reveals that 
a rising trend in the final data points of a continuous observing block is relatively common. 
This phenomenon may be attributed to systematic photometric biases when the source is observed at low elevation, 
as it approaches the limit of ZTF's observable window at the end of a block. 
Nevertheless, a brightening as large as $\sim 0.5 ~\!{\rm mag}$, as seen for J143041.67+355703.8, remains rare. 
Furthermore, based on the spatial localization volume of \ac{GW}190412 
($\bar z \approx 0.15, ~\Delta V_{\rm c} \approx 10^7 ~\!{\rm Mpc}^3$, \citet{2020PhRvD.102d3015A}), 
the average number density of AGNFCC flares 
($\sim 0.4 \times 10^{-7}~\!{\rm Mpc}^{-3}$ per 200 days, 
derived from $\sim 6,400$ flares with $z<0.5$ out of $1.7 \times 10^{10} ~\!{\rm Mpc}^3$ total comoving volume), 
we estimate that the expected number of flares ZTF could detect within the \ac{GW}190412's error volume is $\sim 0.4$. 
This implies that the probability of finding at least one candidate flare from the AGNFCC 
that matches the \ac{GW}190412 skymap by pure chance reaches $\gtrsim 33\%$.

To further investigate whether GW190412 event and J143041.67+355703.8 candidate flare share a common origin, 
we examined the properties of the GW190412 source and the J143041.67+355703.8 candidate flare's host AGN. 
The inferred intrinsic parameters of the GW190412 event, which may indicate its formation channel, 
are as follows 
\citep{2020PhRvD.102d3015A, 2021PhRvX..11b1053A}: 
i) source-frame mass of more-massive \ac{BH} $m_1 = 30.1^{+4.6}_{-5.3} ~\!\!M_{\odot}$, 
ii) mass ratio $q \equiv m_2/m_1 = 0.28^{+0.12}_{-0.07}$, 
iii) dimensionless spin magnitudes of component \acp{BH} $(\chi_1 = 0.44^{+0.16}_{-0.22}, \chi_2 \sim 0.5^{+0.4}_{-0.4})$, 
iv) effective inspiral spin parameter $\chi_{\rm eff} = 0.25^{+0.08}_{-0.11}$, and
v) formation eccentricity $e_{\rm f} \sim 0.34_{-0.29}^{+0.53}$ \citep{2025arXiv251222044B}. 
The extreme mass-ratio, non-zero pre-merger spins and formation eccentricity in GW190412 
suggest that it likely formed through a hierarchical merger channel \citep{2020PhRvL.125j1103G, 
2020ApJ...896L..10R, 2020MNRAS.498..495D, 2020ApJ...901L..39O, 2021MNRAS.502.2049L}. 
Its nonzero effective spin further supports an origin in 
an \ac{AGN} environment \citep{yang19mass-spin, 2020ApJ...903..133S, 2020ApJ...899...26T, 2021ApJ...908..194T, gayathri21}. 

Regarding the host AGN of the J143041.67+355703.8 candidate flare, 
its apparent $B$-band magnitude is recorded as $21.04 ~\mathrm{mag}$ \citep{2023OJAp....6E..49F}. 
Adopting a typical bolometric correction of $L_{\rm bol}/L_B \approx 16$ 
for a bolometric luminosity level of $\sim 10^{43} ~\!\mathrm{erg \cdot s^{-1}}$ \citep{2007ApJ...654..731H, 2020MNRAS.495.3252S}, 
we roughly estimate its bolometric luminosity to be $L_{\rm bol} \approx 3.3 \times 10^{43} ~\!\mathrm{erg \cdot s^{-1}}$. 
Furthermore, based on the host AGN spectrum from the first data release of the Dark Energy Spectroscopic Instrument 
\citep[DESI;][]{2025arXiv250314745D}, we estimate the mass of its central massive \ac{BH} 
to be $M_{\rm MBH} \approx 10^7 ~\!M_{\odot}$ 
(see Appendix \ref{sec:appendix_spectrum} for the spectrum of the J143041.67+355703.8 flare's host AGN). 
From $L_{\rm bol}$ and $M_{\rm MBH}$, the Eddington ratio of 
the host AGN is estimated as $\lg \lambda_{\rm Edd} \approx -1.6$, 
a value lower than that of $70\%$ of AGNs with $L_{\rm bol} < 10^{45} ~\!\mathrm{erg \cdot s^{-1}}$ 
presented in \citet{2022ApJS..263...42W}. 
The relatively lower $L_{\rm bol}$ and lower $\lambda_{\rm Edd}$ of the host AGN of J143041.67+355703.8 
are consistent with the property constraints reported for BBH-host AGNs \citep{Zhu:2025rdv}. 
They also match the characteristic AGN properties predicted by theoretical studies 
\citep[e.g.,][]{2024MNRAS.530.2114G, 2025ApJ...982L..13G} that 
suggest the presence of migration traps for \acp{BH} in AGN disks. 

In summary, the intrinsic parameters of the \ac{GW} and 
the properties of the flare's host AGN 
collectively support an association between the \ac{BBH} merger GW190412 and the candidate flare J143041.67+355703.8. 
We note, however, that the ZTF obtained only two observations during the candidate flare's peak phase. 
We have also searched for the J143041.67+355703.8 flare in data from other transient surveys, 
including the Transient Name Server (TNS), 
the Asteroid Terrestrial-impact Last Alert System \citep[ATLAS;][]{2018PASP..130f4505T}, 
and the All-Sky Automated Survey for Supernovae \citep[ASAS-SN;][]{2023arXiv230403791H}. 
Similarly, no effective observations are available from these archives 
during the candidate flare's peak period.

\section{Discussion}   \label{sec:discussions}

In this work, we estimate the fraction of the LVK \ac{BBH} merger events that are associated with AGN flares observed by ZTF. 
Our preliminary result suggests that $f_{\rm flare} = 0.07_{-0.05}^{+0.24}$ 
of BBH mergers may have an associated ZTF flare. 
However, this non-zero $f_{\rm flare}$ is primarily contributed by the association 
between the GW190412 event and the J143041.67+355703.8 candidate flare. Despite ZTF having only two effective observations 
during the peak period of the J143041.67+355703.8 flare, the intrinsic properties of the GW190412 merger event and 
the properties of the host AGN of the J143041.67+355703.8 candidate flare both support a common origin for them. 

The theoretical interpretation of J143041.67+355703.8 flare as a flare produced by the \ac{GW}190412 merger is straightforward. 
From the observed light-curve amplitude ($\sim 0.5 ~\!{\rm mag}$) of the J143041.67+355703.8 candidate flare 
and its fitted duration ($\sim 27$ days), the total energy released by the J143041.67+355703.8 flare 
is estimated to be $O(10^{49} ~\!{\rm erg})$. An energy release of this magnitude can be explained by 
the same model proposed for a BBH merger–induced flare in \citet{graham20flare}, 
in which the post-merger \ac{BH} drags its bound gas to collide with the surrounding disk gas, 
producing optical thermal emission. In \citet{graham20flare}, the BBH corresponding to \ac{GW}190521 had 
a total mass of $M_{\rm BBH} \sim 160 ~\!M_{\odot}$, and its candidate flare exhibited a duration of $\sim 50$ days 
and a total energy release of $O(10^{51} ~\!{\rm erg})$. 
In the present case, the \ac{GW}190412 event has a total mass of $M_{\rm BBH} \sim 40  ~\!M_{\odot}$. 
Using the radiative luminosity model from \citet{graham20flare} \citep[also see][]{2019ApJ...884L..50M}, 
$L_{\rm bol} \propto \eta \cdot \rho \cdot M_{\rm BBH}^2 \cdot v_k^{-3}$, 
where $\eta$ is the radiative efficiency, $\rho$ is the gas density in the \ac{AGN} disk, 
and $v_k$ is the kick velocity of the BBH, 
the total energy released by J143041.67+355703.8 flare can be explained by 
adopting parameter values similar to those used for the candidate flare associated with GW190521: 
$\eta \sim 0.1$, $\rho \sim 10^{-10} ~\mathrm{g \cdot cm^{-3}}$ and $v_k \sim 200 ~\mathrm{km \cdot s^{-1}}$. 
This kick velocity lies within the constraints from LVK \ac{GW} detections \citep{2022arXiv221103465C, 2025arXiv251222044B}. 

If the J143041.67+355703.8 flare is considered a genuine \ac{EM} counterpart to the GW190412 merger, 
the Hubble-Lema\^itre constant $H_0$ can be constrained using its host's spectroscopic redshift from DESI 
and the \ac{GW} luminosity distance from LVK detection (here, we used the mixed skymap). 
Adopting a uniform prior $H_0 \in [20, 200] ~\!\mathrm{km \cdot s^{-1}\! \cdot Mpc^{-1}}$ and 
a fixed flat-$\Lambda$CDM cosmology with $\Omega_M = 0.308$ in a Bayesian framework \citep[refer to][]{2025ApJ...986...61L}, 
we obtain $H_0 = 82.9_{-14.5}^{+40.8} ~\!\mathrm{km \cdot s^{-1}\! \cdot Mpc^{-1}}$. 
This central value agrees closely with the value derived from \ac{GW}190412 combined with the DESI galaxy catalog \citep{2023RNAAS...7..250B}. 
Such consistency supports, or at a minimum does not contradict, a physical connection between 
the J143041.67+355703.8 flare and the \ac{GW}190412 merger. 

Furthermore, it is worth noting that this work and \citet{Zhu:2025rdv} employed the same statistical framework 
to test the spatial correlation between \acp{BBH} and \acp{AGN}, with the key distinction being that 
this work specifically targeted \acp{AGN} exhibiting temporally coincident flare activity. 
The GW190412 event, which provides the dominant support for the non-zero most probable $f_{\rm flare}$ in this work, 
was also included in the analysis of \citet{Zhu:2025rdv}. 
However, its contribution to favoring a non-zero fraction parameter $f_{\rm agn}$ in that work was not significant. 
This discrepancy primarily arises because the host AGN of the candidate flare counterpart J143041.67+355703.8 
associated with GW190412 in this work was identified from the \textsf{Milliquas} \citep{2023OJAp....6E..49F} 
and \textsf{DESI} \citep{2025arXiv250314745D} \ac{AGN} catalogs, but was not recorded in 
the Sloan Digital Sky Survey (SDSS) AGN catalog (see \citet{2020ApJS..250....8L} or \citet{2022ApJS..263...42W}) 
used by \citet{Zhu:2025rdv}. 

In conclusion, our search for candidate flares  associated with the GWTC-4.0 BBH events, 
together with statistical constraints on the flare fraction parameter $f_{\rm flare}$, 
provides preliminary indications that a fraction of BBH mergers may be linked to ZTF flares. 
We caution that the non-zero estimate of $f_{\rm flare}$ is driven primarily by GW190412. 
Furthermore, the light curves of GW190412's sole candidate flare contain merely a single data point 
in each of the $g$ and $r$ bands during the peak period, which does not meet the criteria for inclusion 
in the refined \ac{AGN} flare sample (such as AGNFRC) and thus can only be regarded as a candidate. 
However, within the context of current non-detections of counterparts \ac{EM} to BBH mergers 
\citep[e.g.,][also see the ``{LVK GW}'' page in TNS for centralized reporting, 
\url{https://www.wis-tns.org/ligo/events}]{2021ApJ...916...47K, 2021MNRAS.508.3910C, 2022ApJ...933...85A, 
2023MNRAS.525.4065B, 2024AnP...53600215P, 2024ApJ...976..123W, 2025PhRvD.112f3019D, 2025arXiv250800291Z, 2026ApJS..282...13H}, 
these preliminary indications strengthen the motivation for continued counterpart searches and 
offer a reference for future search strategies. 
Clearly, establishing definitive associations -- both statistical and individual -- between BBH events 
and candidate flares remains a key objective. With the ongoing operations of the LVK network \citep{LVK_plans}, 
and the advent of new time-domain surveys such as 
the Wide Field Survey Telescope \citep[WFST;][]{2023SCPMA..6609512W, 2024ApJ...964L..22H}, 
the Vera C. Rubin Legacy Survey of Space and Time \citep[LSST;][]{2009arXiv0912.0201L, 2023PASP..135j5002H}, 
and the Einstein Probe \citep[EP;][]{2025SCPMA..6839501Y}, 
it is expected that the connection between BBH mergers and their potential counterparts \ac{EM} will become increasingly clear.

\section*{Acknowledgements}
The authors thank Zheng-Yan Liu, Shifeng Huang, Yibo Wang, Ken Chen, Rui Niu, Yin-Jie Li, Tong Chen, Guo-Peng Li and Chengjie Fu 
for helpful discussions. 
This work is supported by the National Natural Science
Foundation of China (grant Nos. 12325301, 12273035, 12473037), Strategic Priority Research Program of the
Chinese Academy of Science (grant No. XDB0550300), the
National Key R\&D Program of China (grant Nos.
2021YFC2203102, 2022YFC2204602, 2024YFC2207500), the Science Research Grants from
the China Manned Space Project (grant No. CMS-CSST-2021-
B01), the 111 Project for ``Observational and Theoretical Research on Dark Matter and Dark Energy'' (grant No.
B23042), Cyrus Chun Ying Tang Foundations and Guizhou Provincial Major Scientific and Technological Program XKBF (2025) 011.

\software{\textsf{numpy} \citep{2011CSE....13b..22V}, 
\textsf{scipy} \citep{2020NatMe..17..261V}, 
\textsf{ligo.skymap} \citep{ligo_skymap}, 
\textsf{LALSuite} \citep{ligo_lalsuite}, 
\textsf{astropy} \citep{2013A&A...558A..33A, 2018AJ....156..123A, 2022ApJ...935..167A}, 
\textsf{alphashape} \citep{alphashape_python}, 
\textsf{pyqsofit} \citep{2018ascl.soft09008G, 2019ApJS..241...34S, 2024ApJ...974..153R}, 
\textsf{matplotlib} \citep{2007CSE.....9...90H} and 
\textsf{seaborn} \citep{Waskom2021}. 
}
\appendix
\section{Spectrum of the host AGN of J143041.67+355703.8}   \label{sec:appendix_spectrum}

\begin{figure*}[tbp]
 \centering
 \includegraphics[width=0.9\textwidth]{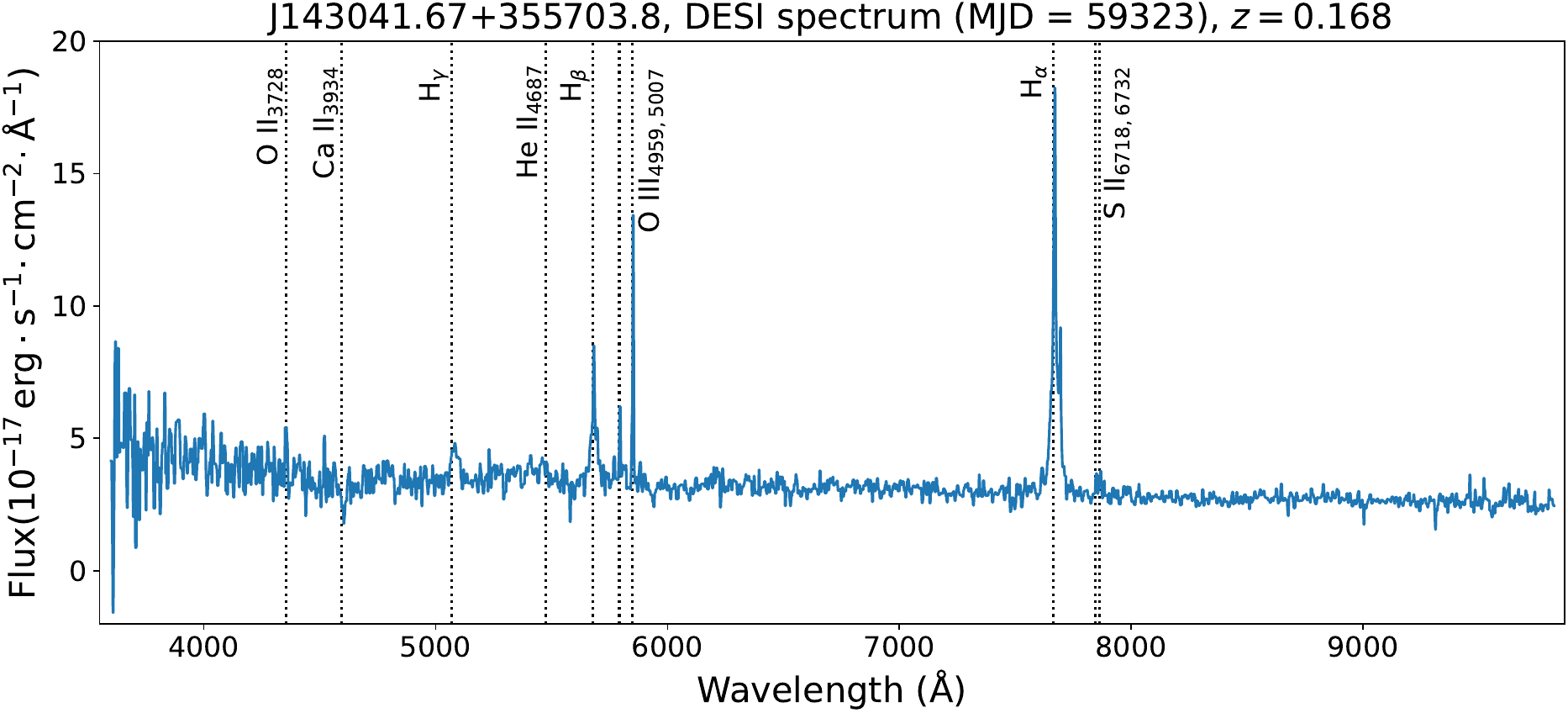}
 \caption{DESI spectrum of the host AGN of J143041.67+355703.8 flare. 
 Vertical dotted lines mark the positions of characteristic AGN emission or absorption lines. 
}
 \label{fig:agn_spectrum}
\end{figure*}

In this appendix, we present the observer-frame DESI spectrum of the host AGN of 
J143041.67+355703.8 \citep{2025arXiv250314745D} (the sole candidate flare for the GW190412 merger)
in Figure \ref{fig:agn_spectrum}. 
Vertical dotted lines mark the positions of its characteristic emission or absorption lines. 
The central massive black hole mass  is estimated from the width of the AGN's broad emission lines, 
combined with the empirical radius–luminosity relation of the broad-line region. 
In this work, we derive the massive black hole mass using the radius-luminosity relation at $5100 ~\!{\rm \AA}$. 

\end{CJK*}

\bibliography{references_SBBHflare}
\end{document}